\documentstyle[12pt,epsfig,here,a4]{article}
\begin{document}
{\bf DESY 99-18}

\vspace*{1.4cm}

\begin{center}

{\Large \bf In-situ measurements of optical parameters}

{\Large \bf in Lake Baikal 
with the help of a Neutrino Telescope}

\end{center}

\vspace{0.3cm}


V.A.Balkanov$^2$, I.A.Belolaptikov$^7$, L.B.Bezrukov$^1$,
N.M.Budnev$^2$,
A.G.Chensky$^2$, \\ I.A.Danilchenko$^1$, Zh.-A.Djilkibaev$^1$,
G.V.Domogatsky$^1$, A.A.Doroshenko$^1$, \\ S.V.Fialkovsky$^4$,
O.N.Gaponenko$^1$,
A.A.Garus$^1$, T.I.Gress$^2$, A.Karle$^{8,*}$,
A.M.Klabukov$^1$, \\ A.I.Klimov$^6$,
S.I.Klimushin$^1$, A.P.Koshechkin$^1$, V.F.Kulepov$^4$,
L.A.Kuzmichev$^3$,  
S.Lovzov$^2$,\\  B.K.Lubsandorzhiev$^1$, 
T.Mikolajski$^8$, M.B.Milenin$^4$,
R.R.Mirgazov$^2$,
A.V.Moroz$^2$, \\
N.I.Moseiko$^3$,  S.A.Nikiforov$^2$, E.A.Osipova$^3$,
D.Pandel$^{8,**}$, A.I.Panfilov$^1$, Yu.V.Parfenov$^2$,\\ 
A.A.Pavlov$^2$,   D.P.Petukhov$^1$,
P.G.Pokhil$^1$, P.A.Pokolev$^2$, E.G.Popova$^3$, M.I.Rozanov$^5$,\\ 
V.Yu.Rubzov$^2$,  I.A.Sokalski$^1$, Ch.Spiering$^8$, O.Streicher$^8$, 
B.A.Tarashansky$^2$, T.Thon$^8$, \\ R.Wischnewski$^8$, I.V.Yashin$^3$

{\it 1 - Institute  for  Nuclear  Research,  Russian  Academy of
  Science
(Moscow),\\ \mbox{2 - Irkutsk} State University (Irkutsk), \\ \mbox{3 -
  Moscow}
State University (Moscow), \\ \mbox{4 - Nizhni}  Novgorod  State
Technical
University  (Nizhni   Novgorod),\\  5 - St.Petersburg State Marine
Technical  University (St. Petersburg), \\ \mbox{6 - Kurchatov} Institute
(Moscow), \\ \mbox{7 - Joint} Institute for Nuclear Research (Dubna), \\
8 - DESY-Zeuthen (Zeuthen) \\
** - now at University of Wisconsin (USA) \hspace{0.5cm}
** - now at University of Irvine (USA)}

\vspace{1.3cm}

{\large \bf Abstract}

We present results of an experiment performed in
Lake Baikal at a depth of about 1 km. The
photomultipliers of an underwater
neutrino telescope under construction at this site
have been
illuminated by a distant laser. The experiment not
only provided a useful cross-check of the time
calibration of the detector, but also allowed
to determine inherent optical parameters of the
water in a way complementary to standard methods.
In 1997, we have measured an absorption length of 22\,m
and an asymptotic attenuation length of 18\,m.
The effective scattering length was measured as
480\,m. Using $\langle \cos \theta \rangle = 0.95 (0.90)$ for the 
average scattering angle, this corresponds to a geometrical 
scattering length of 24 (48)\,m.

\vspace{5mm}
\begin{center}
{\it submitted to APPLIED OPTICS} 
\end{center}

\newpage

\section{Introduction}

Propagation of light in optical media is governed by two basic phenomena:
absorption and scattering. In the first case the photon is lost, in the second
case it changes its direction. The inherent parameters generally chosen as a
measure for these phenomena are the absorption length $\lambda _{abs}$, the
scattering length $\lambda _{sct}$ and the scattering function (or
scattering tensor if polarization is taken into account) $\beta \left(
\theta \right) $.

The determination of optical parameters for the water of deep lakes or
oceans meets considerable problems. Typical values for $\lambda _{abs}$ in
the window of maximum transparency are about 20\,m for clearest
 water in deep lakes \cite{Bezr90a,Bezr90b,Gapo96} and 
50\,m  for deep oceans \cite{Zane81,Brad84,Ann95}. Another
important property of natural water is the strongly forward peaked scattering,
with $\left\langle \,\cos \ \theta \,\right\rangle$ = 0.85\,--\,0.96 in oceans
\cite{Jerl76} or lakes \cite{Gapo96,Gapo94,Tara94} and
scattering lengths of 10\,--\,50\,m. 
Because of the long absorption
and scattering lengths and the steep scattering function, devices to measure
these optical parameters in clear water have to be extremely precise and
well calibrated.

{\it In vitro} measurements suffer from the comparatively short base length
of the laboratory devices and from the fact that the samples of the medium
can deteriorate before being analyzed in laboratory. Furthermore, for {\it in
vitro} measurements it is rather difficult to perform long-term experiments
and to observe seasonal variations of the optical parameters. {\it In situ}
devices are difficult to handle, particularly for
greater depths. On the other hand, only {\it in situ} measurements allow
long term observations of the medium, including the implications of
temporary changing mechanical, chemical, biological and thermal
conditions.

In the course of preparing a deep-underwater 
neutrino telescope in Lake Baikal,
{\it in situ} optical measurements have been carried out 
over several years. 
We have constructed a special device (see \cite{Bezr90a,Tara95})
and developed proper methods  \cite{Gapo96,Gapo94,Tara94}
to determine and to monitor the hydro-optical parameters of the water body
at the site of Baikal Neutrino Telescope.  The data obtained from
these measurements do not only yield necessary input parameters
for the operation of the telescope, but are interesting as well
for environmental science.

In this paper we present the results of optical measurements obtained 
in a way complementary to standard limnological (oceanological)
methods, using a laser as light source and
an underwater neutrino telescope as light detector. This
kind of telescope represents a new class of giant  
instruments coming into
operation at various locations in deep oceans and lakes \cite{Review},
and even in glaciers \cite{Amanda}. Being
permanently operational over several years, underwater telescopes
can perform, as a
by-product and for calibration purposes, measurements of optical parameters
as well as of water currents, bioluminescence etc.

We have measured the spatial and temporal distribution
of monochromatic light from a pulsed 
isotropic, point-like source. From the
analysis of the spatial distribution, we derive the values for
absorption length $\lambda _{abs}$ and for the so-called asymptotic 
attenuation length $%
\lambda _{asm}$. 
From the time delay of the
short light flashes arriving at the photomultipliers of the telescope,
we derive $\lambda _{eff}$, which is an effective parameter 
describing scattering.

We describe the Baikal Neutrino Telescope in section 2, and the
laser experiments to determine the optical parameters of the water
in section 3. In section 4 we derive the expressions for the
light field as a function of the distance to an isotropic
source. The experimental data are analyzed in section 5, and values 
for parameters characterizing absorption 
and scattering are derived. Section 6 contains the conclusion.

\section{The Baikal Neutrino Telescope}

The Baikal Neutrino Telescope \cite{APP,Proj}
exploits
the deep water  of the Siberian lake as a detection medium 
for secondary charged particles, like muons or electrons,
generated in neutrino interactions. 
Neutrinos are etheric particles characterized by
their extremely rare interaction with all kind of matter.
They are supposed to be generated in cosmic particle
accelerators like the nuclei of Active Galaxies or
binary star systems. Shielded by kilometers of
water burden, underwater telescopes aim to detect
the rare reactions of neutrinos from these exotic sources.
 
A lattice of photomultiplier tubes (PMTs) spread
over a large volume records the Cherenkov light emitted 
by the relativistic charged particles. The light 
arrival times at the locations of the
PMTs are measured with an accuracy of
a few nanoseconds. From the time pattern, the
particle trajectory can be reconstructed with a
resolution of 1\,--\,2 degrees. In addition to the
arrival times, the amplitudes of the
signals are recorded. Typical amplitudes for
particles triggering the array are in the
few-photoelectron range.

\begin{figure}[H]
\centering
  \mbox{\epsfig{file=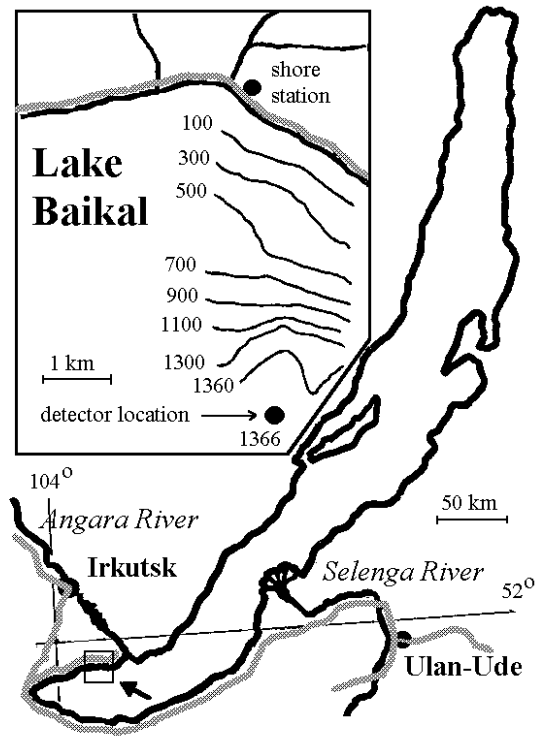,height=10.5cm}}
\caption[4]{\small
Location of the Baikal experiment
}
\end{figure}

The Neutrino Telescope {\it NT-200} is being deployed in the
southern part of Lake Baikal (fig.1). The distance
to shore is 3.6 km, the depth of the lake is 1366 m at this
location. Fig.2 sketches the instrumentation of the site.
Strings anchored by weights at the bottom are held in vertical
position by buoys at various depths. The deployment of the
detector elements is carried out in late winter, when the 
lake is covered by a thick layer of ice. Three cables
({\it 1,2,3} in fig.2) connect the detector with the shore
station. Each shore cable ends at the top of a string
({\it 4,5,6}, respectively). String {\it 7} carries the 
telescope. A special "hydrometric" string {\it 8} 
is equipped with instruments to measure the optical
parameters of the water as well as water currents, temperature,
pressure and sound velocity. The spatial coordinates of the 
components of the telescope are monitored by an ultrasonic 
system consisting of transceivers {\it 9-14}, and receivers
along the strings. The relative coordinates of the
components are determined with an accuracy of about 20 cm.

\begin{figure}[H]
\centering
  \mbox{\epsfig{file=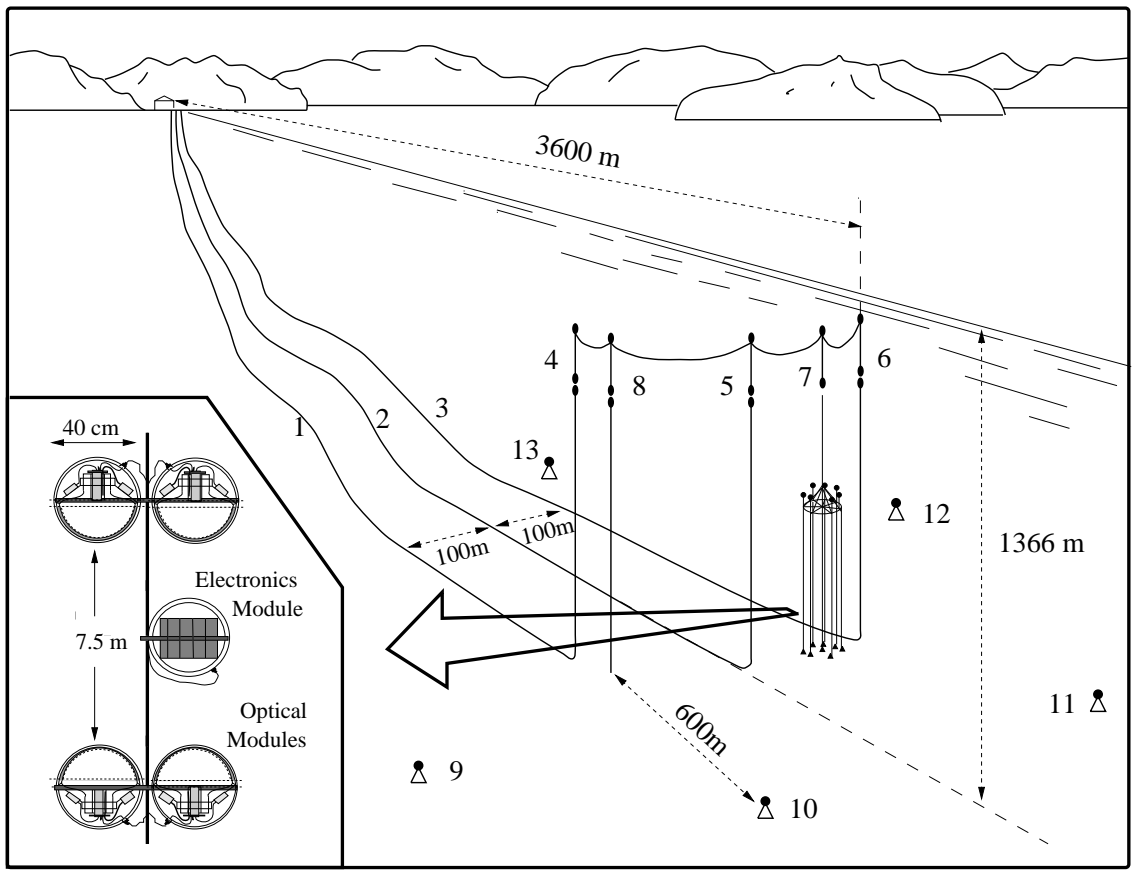,height=15cm,angle=-90}}
\caption[4]{\small
Overall view of the {\it NT-200} complex. 
                   {\it 1,2,3} -- cables to shore,
                  {\it 4,5,6} -- string stations for shore cables
                  {\it 7} -- string with the telescope,
                  {\it 8} -- hydrometric string,
                  {\it 9-14}  -- ultrasonic emitters.
The detail bottom left shows two pairs of optical modules (OMs), together
with the electronics module controlling the OMs. Shown are two
pairs directed face to face.
}
\end{figure}

The telescope {\it NT-200} consists of 8 sub-strings arranged at the
center and the edges of an equilateral heptagon. The 
light detection elements are optical
modules (OMs) containing a hybrid phototube {\it Quasar-370}
with a hemispherical photocathode of 370 mm diameter. 
The time resolution of the {\it Quasar-370} is 3 nsec for
a single-photoelectron signal and improves to about
1 nsec for large amplitudes. The
OMs are arranged pairwise, with the two phototubes being
operated in coincidence (see fig.2).
If an event triggers at least
3 pairs within 500 nsec,
light arrival time and amplitude 
of each hit pair ({\it "channel"}) are recorded.
Each of the 8 strings of NT-200 will carry 24 OMs.

The time calibration is performed with a help of a nitrogen laser
positioned just above the array \cite{Thomas}. 
The 1-nsec light flashes of
the laser are transmitted by optical fibers of equal length
to each of the OM pairs.

\section{The Laser Experiments}

For the two experiments described in this paper, an additional
laser device was operated  
at various locations with respect to the PMTs of the 
neutrino telescope. Short light
pulses were emitted by the laser, traveled through the water,
 and were detected by the PMTs
during normal operation,
when the standard trigger on muons crossing the array
(3-fold coincidence within 500\,ns) was set.

The laser module \cite{Thomas}
contains a pulsed nitrogen laser (wavelength 337.1\,nm) 
which pumps a dye laser with an emission maximum
at 475\,nm.  This wavelength well matches the point of maximum
transparency at 490\,nm, where Baikal water has an absorption length
of about 20\,m.

\begin{figure}[H]
\centering
  \mbox{\epsfig{file=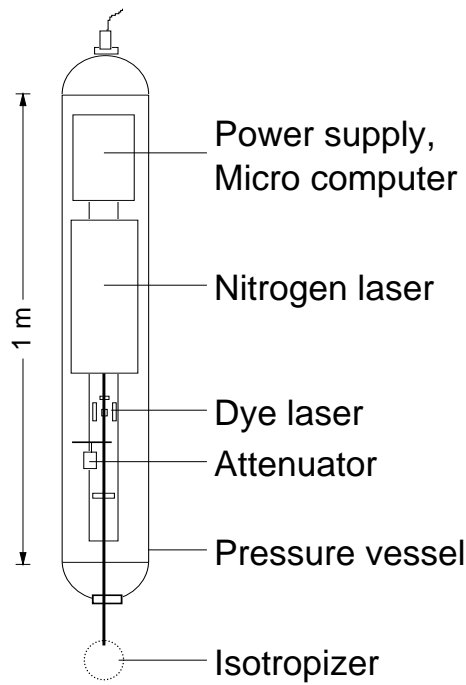,height=9cm}}
\caption[4]{\small
Schematic view of the laser module
}
\end{figure}

The light pulses generated by the laser have a
length of about 0.5\,ns (FWHM), thus being shorter than the time
resolution of the photomultipliers.  An attenuation disk, moved by a
stepper motor, is used to attenuate the light beam in five steps from
100\,\% to 0.3\,\% of the laser intensity of about $10^{12}$ photons per
pulse.  Both lasers, the attenuator, power supplies and a 
micro-computer are contained in a glass pressure vessel 
\cite{Nautilus} of 155\,mm
diameter and 1\,m length (see fig.3).  After having crossed the vessel
wall, the laser beam is directed into a hollow sphere with a
diffusely reflecting inner surface.  Through small holes in the sphere
the light is able to escape, with a nearly isotropic distribution.
Due to the large distance to the detector, the laser 
module can be considered as an isotropic point-like light source.

The laser can be triggered to generate cycles of light
pulses. Each cycle consists of 5 series of 200 equally intense
pulses. The intensities of consecutive series
differ  by a factor of 3 to 4.
Laser-induced events fulfill
the trigger conditions for muons crossing the array.
They can be separated from 
those using the periodicity of
the laser pulses (9.1\,Hz). The remaining background
from muon events ranges from  $10^{-3}$ for low light intensities
to  $10^{-4}$ for higher intensities.

In the following, data taken with 2 configurations are analyzed:

\begin{itemize}

\item {\it 1995 experiment}

In April 1995, an intermediate version of the final array
{\it NT-200} was deployed. It carried 72 optical modules mounted
at four short and one long string (see fig.4). The laser experiment
has been performed immediately after
deployment of this detector ({\it NT-72}).
The laser module was deployed
from the ice surface and
placed at several locations 20 to 200\,m away from the
PMTs.  Preliminary results from this experiment have
 been published in \cite{Albrecht}.

\begin{figure}[H]
\centering
  \mbox{\epsfig{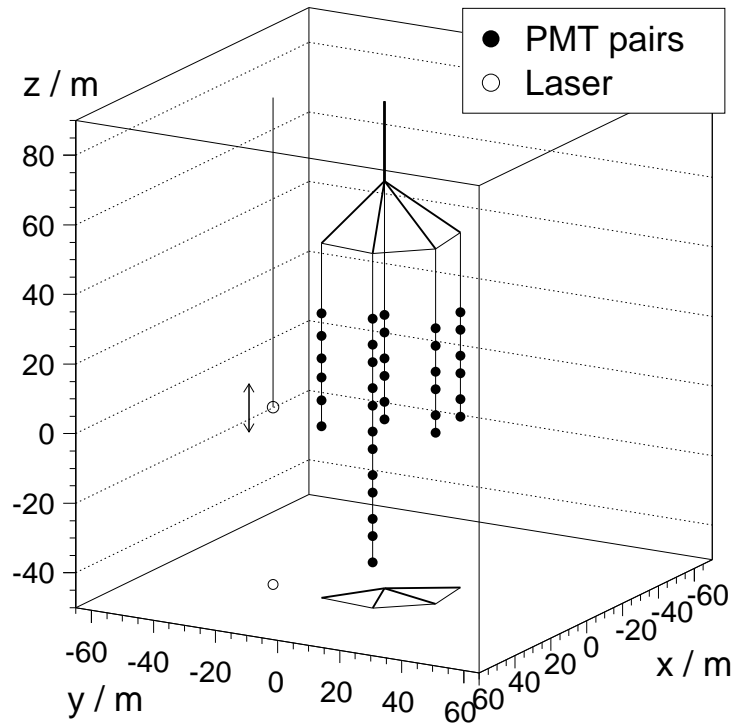}}
\caption[4]{\small
Schematic view of the 1995 configuration
of the laser experiment.
Each filled circle represents a pair of optical modules (channel).
The laser is indicated by the open circle.
}
\end{figure}

\newpage

\item{\it 1997 experiment}

The 1997 laser experiment was performed after having
deployed the first string of the 1997 season.
The string carried 24 PMTs, with the highest at 1059 m depth and
the lowest at 1128 m depth.
The laser was moved along the string at a horizontal distance
of 12 meters to the string with the PMTs, starting at a
height close to the top PMT. The last of the 15 positions
was 221 m below the first, at 1286\,m (see fig.5).

\end{itemize}

In both experiments, the times and amplitudes of the
fired channels were recorded if the 
standard trigger on muons crossing the array
(3-fold coincidence within 500\,ns) was fulfilled.

\begin{figure}[H]
\centering
  \mbox{\epsfig{file=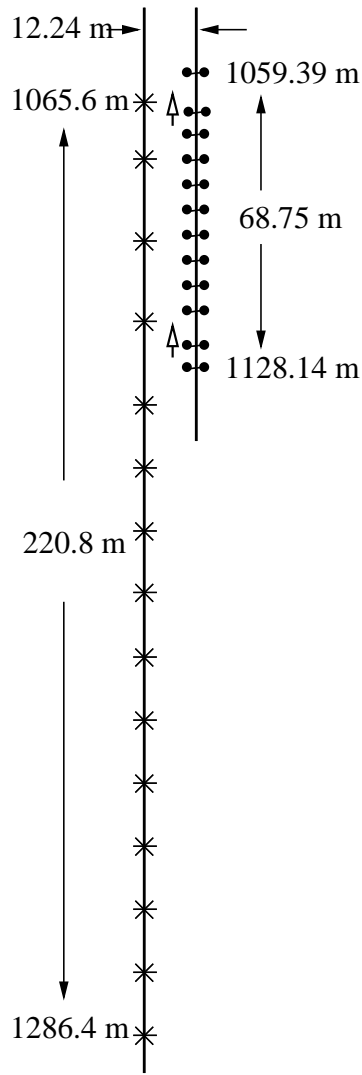,height=3.5cm,angle=-90}}
\caption[4]{\small
Schematic view of the the 1997 configuration of the laser
experiment. Channels are shown explicitly as pairs of OMs.
The various positions of the laser are indicated by stars.
The OMs face down, with the exception of channels
2 and 11 (upward arrows) which are directed upward.
}
\end{figure}

\newpage

\section{The light field of an isotropic source in a medium}

The photon field $\Phi \left( R,\,\vec \Omega ,\,t\right) $ produced in a
medium by a pulsed point-like source of monochromatic light is a function of
distance $R$ from the light source, direction vector $\vec \Omega $, and
time $t$ after emission. This function can be calculated
by Monte Carlo methods, with $\lambda _{abs}$, $%
\lambda _{sct}$ and $\beta \left( \theta \right) $ 
as input parameters.  In this section we derive analytical
approximations for the photon field and compare it to the results
of Monte Carlo simulations.

In the special
case of a medium without scattering one has

\begin{equation}
\label{3.a}\Phi \left( R,\,\vec \Omega ,\,t\right) =F\left( R\right) \delta
\left( \vec \Omega \right) \delta \left( t-R/v\right) ,
\end{equation}

where $F(R)$ is the photon flux:

\begin{equation}
\label{3.b}F\left( R\right) =\int \,\Phi \left( R,\,\vec \Omega ,\,t\right)
\,d\vec \Omega \,dt=\frac{I_0}{4\pi R^2}\,e^{-\frac R{\lambda _{abs}}}.
\end{equation}

For a medium with scattering, $F\left( R\right) $ can be
written as

\begin{equation}
\label{3.1}F\left( R\right) =\frac{I_0}{4\pi R^2}\,e^{-\frac R{\lambda
_{abs}}\,\mu \left( R\right) },
\end{equation}

where, in the general case, $\mu \left( R\right) $ is some intricate
function of $R$. This function takes into account the increase of the
photon path due to scattering. 
It was shown in \cite{Bauer} that $1<\mu \left( R\right) <1.01$
for distances $R<0.6\,\lambda_{sct}$ and
steep scattering functions ($\left\langle \cos \,\theta
\right\rangle \geq 0.9$). 
At larger distances, however, the contribution from
scattering becomes  significant. In the following
we estimate the influence of scattering  on the photon flux at
moderate distances from the source. 

Due to scattering, a photon does not
travel along a straight line but follows some polygonal path with random
vectors $\vec r_i$ ($i=1,\ 2,\ ...\,n$) which are the trajectories of a
photon between two successive scattering acts. Thus,

\begin{equation}
\label{3.2}\left\langle r_i\right\rangle =\lambda _{sct},
\end{equation}

\begin{equation}
\label{3.3}\left\langle \vec r_i\cdot \vec r_{i+1}\right\rangle =\lambda
_{sct}^2\,\left\langle \cos \,\theta \right\rangle .
\end{equation}

It turns out that under rather general assumptions about the shape of the
scattering function, the following relation takes place:

\begin{equation}
\label{3.4}\left\langle \cos \,\theta _{i\,,\,i+k}\right\rangle
=\left\langle \cos \,\theta \right\rangle ^k,
\end{equation}

where $\theta _{i\,,\,i+k}$ is the angle between the vectors $\vec r_i$ and $%
\vec r_{i+k}$. In this case

\begin{equation}
\label{3.5}R=\sqrt{\sum\limits_{i=1}^n\sum\limits_{j=1}^n\left\langle \vec
r_i\cdot \vec r_j\right\rangle }=\lambda _{sct}\sqrt{n+2\sum\limits_{i=1}^n%
\left( n-k\right) \left\langle \cos \,\theta \right\rangle ^k}.
\end{equation}

The average length $L$ of the photon path for $n$ successive scattering acts
is

\begin{equation}
\label{3.6}L=n\cdot \lambda _{sct}.
\end{equation}

For simplicity we consider the case of $\left\langle \cos
\,\theta \right\rangle$ close to 1. 
With $\left( 1-\left\langle \cos \,\theta
\right\rangle \right) $ being small, from
from eqs.(\ref{3.5}) and (\ref{3.6}) one gets

\begin{equation}
\label{3.7}L\left( R\right) \approx R\sqrt{1+\frac 13\frac R{\lambda
_{sct}}\,\left( 1-\left\langle \cos \,\theta \right\rangle \right) }.
\end{equation}

For water with strongly forward peaked scattering we finally obtain

\begin{equation}
\label{3.8}F\left( R\right) =\frac{I_0}{4\pi R^2}\,e^{-\frac R{\lambda
_{abs}}\,\sqrt{1+\frac 13\frac R{\lambda _{sct}}\,\left( 1-\left\langle \cos
\,\theta \right\rangle \right) }}.
\end{equation}

We note that in this equation 
$\lambda _{sct}$ and $\left\langle \cos \,\theta
\right\rangle $ are combined in an expression
$\lambda_{sct}/(1 - \langle \cos \theta \rangle)$,
resulting in a reduction of the 
number of independent parameters in eq. (\ref
{3.8}) from three to two. This agrees with the conclusions
drawn in \cite{Pelew}  from the measurements of the photon
flux of an isotropic source in artificial media 
as well as in natural water.
For all these strongly forward scattering media it was found 
that $F\left( R\right) $ can be
described as a function of only two parameters, $\lambda _{abs}$ and
$P$, where

\begin{equation}
\label{3.9}P=\frac{\lambda _{abs}}{\lambda _{sct}}\,\frac{\left\langle
\theta ^2\right\rangle }2.
\end{equation}

Since for $\left\langle \cos \,\theta \right\rangle \simeq 1$ one has $%
\left\langle \theta ^2\right\rangle \approx 2\left( 1-\left\langle \cos
\,\theta \right\rangle \right) $, eq.(\ref{3.8}) can be rewritten in terms
of the parameters $\lambda _{abs}$ and $P$ (see also
\cite{Dobrynin,Dolin1}):

\begin{equation}
\label{3.10}F\left( R\right) =\frac{I_0}{4\pi R^2}\,e^{-\frac R{\lambda
_{abs}}\,\sqrt{1+\frac 13\frac R{\lambda _{abs}}\,P}}.
\end{equation}

Our Monte Carlo simulations  \cite{Lanin} 
confirm the results
of \cite{Pelew} which proved for an interval 
$\langle \theta^2 \rangle$ = 0.114 to 0.582
and for distances $R < 5\,\lambda _{abs}$
that the error in $F(R)$ resulting from the reduction of
$\lambda_{scatt}$ and $\beta(\theta)$ to $P$
is smaller than than $1\%$.

With the notation

\begin{equation}
\label{3.11}\lambda _{eff}=\frac{\lambda _{sct}}{1-\left\langle \cos
\,\theta \right\rangle },
\end{equation}

eq.(\ref{3.8}) can be written as

\begin{equation}
\label{3.12}F\left( R\right) =\frac{I_0}{4\pi R^2}\,e^{-\frac R{\lambda
_{abs}}\,\sqrt{1+\frac 13\frac R{\lambda _{eff}}}}.
\end{equation}

The definition of $\lambda _{eff}$ according to eq.(\ref{3.11}) coincides
with the one for the diffusion case provided $\left\langle \cos \,\theta
\right\rangle $ is close to zero (i.e. quasi-isotropic scattering)
and in the expansion in
powers of $\left\langle \cos \,\theta \right\rangle $ only the first
correction term is taken into account. We  emphasize 
that  
in our case $\lambda _{eff}$ is just a convenient, artificial parameter
(comparable, e.g., to the parameter $P$ in eq. (\ref{3.10})) and should not
be associated with an ''effective'' increase of the scattering length due to
the anisotropy of the scattering.

Although eq.(\ref{3.12}) was derived here for strongly forward
peaked scattering functions, similar considerations can be also performed
for the case where $\left\langle \cos \,\theta \right\rangle $ is not close
to unity as, for instance, in glacier ice which is exploited for the {\it %
AMANDA} neutrino telescope \cite{Amanda,Ice}.

Eq.(\ref{3.10}) 
may also be considered as a particular solution of the transport equation for
the small-angle approximation. In the framework of the small-angle
approximation the general solution has the form (see e.g. \cite
{Monin}):

\begin{equation}
\label{3.13}F\left( R\right) =\frac{I_0}{4\pi R^2}\,\frac{\left( R\,\sqrt{P}%
/\lambda _{abs}\right) }{\sinh \left( R\,\sqrt{P}/\lambda _{abs}\right) }%
\,e^{-\frac R{\lambda _{abs}}}.
\end{equation}

Eq.(\ref{3.10}) can be formally found from the solution (\ref{3.13}) by
expansion with respect to the small parameter $R\,\sqrt{P}/\lambda _{abs}$. The
opposite case, $ R\,\sqrt{P}/\lambda _{abs} \gg 1$, gives

\begin{equation}
\label{3.14}
\begin{array}{c}
F\left( R\right) =
\frac{I_0\sqrt{P}}{2\pi R\lambda _{abs}}\,e^{-\frac R{\lambda _{abs}}\left(
1+\sqrt{P}\right) }= \\  \\
=\frac{I_0}{2\pi R\sqrt{\lambda _{abs}\lambda _{eff}}}\,e^{-\frac R{\lambda
_{abs}}\left( 1+\sqrt{\lambda _{abs}/\lambda _{eff}}\right) }.
\end{array}
\end{equation}

In fig.6 we compare the results of our Monte Carlo simulations
(histogram) with the predictions of eq.(\ref{3.10}) (dashed line) and 
eq.(\ref{3.13}) (solid line) for $\lambda _{abs}=23$\,m and $P=0.1$. The
dotted line represents the result of eq.(\ref{3.1}) with $%
\mu \left( R\right) =1$ (the case without scattering).
One observes a perfect agreement of the approximation (\ref{3.13})
with the Monte Carlo calculations over a wide region of distances. 
For not too large distances also eq.(\ref{3.10}) agrees with the MC
data.

In next section we will apply the expressions obtained to the experimental
data.

\begin{figure}[H]
\centering
  \mbox{\epsfig{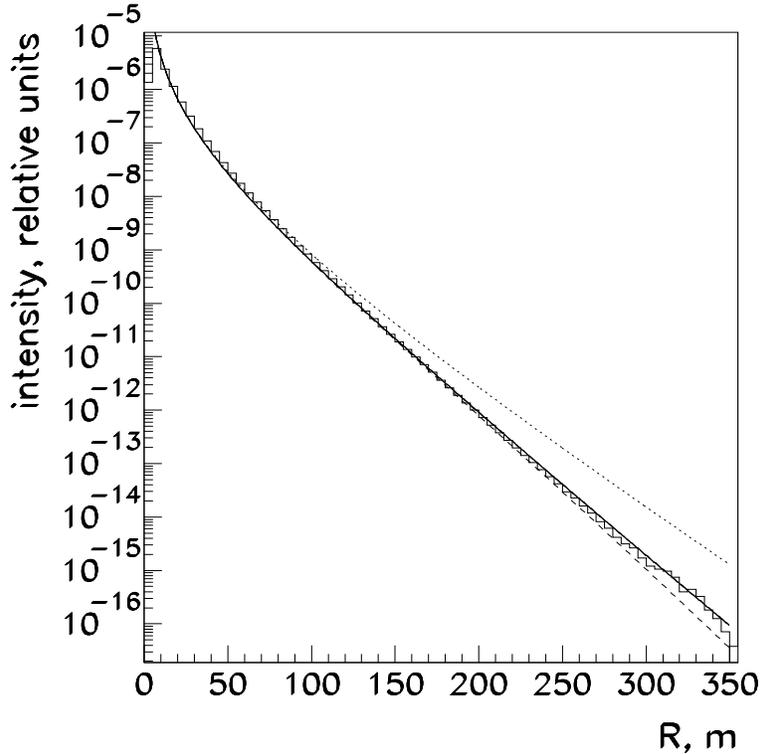}}
\caption[4]{\small
Intensity vs. distance as obtained from Monte Carlo calculations
(histogram), eq.(\ref{3.10}) (dashed line) and eq.(\ref
{3.13}) (solid line). In all three cases $\lambda_{abs}$=23\,m 
and $P=0.1$ have been used. The dotted
line is the result of eq.(\ref{3.1})
with $\mu(R) = 1$ 
}
\end{figure}

\section{Analysis of experimental data}

\subsection{Absorption}

{\large \it 1995 experiment}

The absorption of light on its way to the
photomultipliers can be determined from the measured
amplitudes. For the 1995 laser experiment,
24 of the 72/2 = 36 channels have been included in the analysis.
The combination of amplitudes from different photomultipliers
requests a careful calibration. After amplitude calibration,  and
using the well-known angular acceptance of the
photomultipliers, the mean amplitudes measured by each channel of 
{\it NT-72} have
been converted to light intensities. In fig.7 the dependence of
the photon density on the laser distance is shown. 
The points represent the
measured signals from the optical modules facing toward the laser, 
for different laser
positions and intensities. The data are well described by the relation

\begin{equation}
\label{4.1}F(R)\sim \frac 1{R^2}\,e^{-\frac R\lambda }
\end{equation}

over a wide region of distances $R$. This is the approximation of
equation(\ref{3.12}) for the case $R/\left( 3\lambda _{eff}\right) \ll 1$.
With $\lambda_{eff} \approx 480$\,m (see sect.5.2)\footnote{Actually,
from the rather precise description of the
data by an $1/R^2 \cdot exp(-R/\lambda)$ dependence, a lower
limit of about 300\,m for $\lambda_{eff}$ can be estimated.} 
and
$R$ covering the range 30\,--\,150\,m, the latter condition
is satisfied. Therefore, $\lambda$ in eq.(\ref{4.1}) 
can be identified with the absorption
length $\lambda _{abs}$. The value of $\lambda _{abs}$ obtained from the
experimental data of 1995 is $(19.9 \pm 0.2$)\,m \cite{Albrecht}.

\begin{figure}[H]
\centering
  \mbox{\epsfig{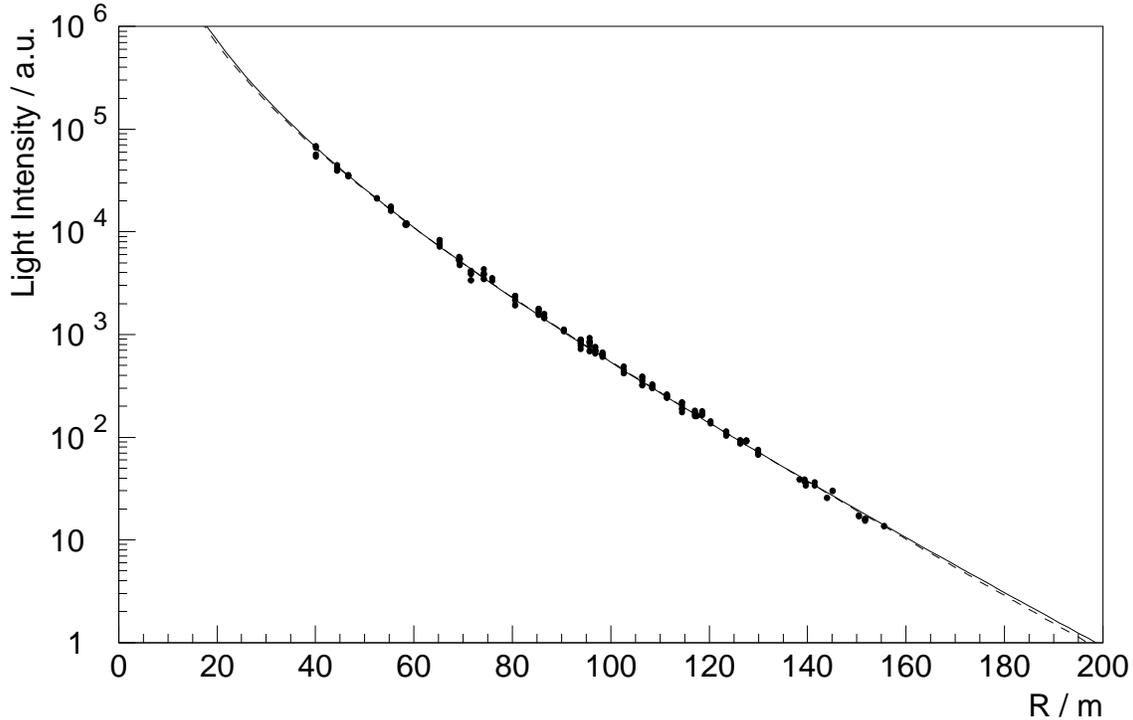}}
\caption[4]{\small
Laser experiment 1995: Light intensity as a function of
distance to the laser. The dots show the measured values, the
solid line shows an exponential decrease with $\lambda$ =
19.9\,m. If scattering is included, the
dependence shown by the dashed line is expected.
}
\end{figure}

\bigskip

{\large \it 1997 experiment}

In fig.8 we present the average amplitudes multiplied with $R^2$
vs. distances $R$, for one of
the channels of the first 1997 telescope string. The
data points can be grouped along five straight lines,
each corresponding to a given intensity of the laser. Contrary to the
experiment of 1995, the analysis has been performed
for each channel separately. This procedure avoids biases
due to possible errors in the relative amplitude calibration
of the PMTs.
The large number of channels (i.e the multiple independent
determination of $\lambda_{abs}$) and the high precision
of the measured amplitudes and
distances results in small errors of $\lambda _{abs}$.
The absorption length is 
$\lambda _{abs}=22$\,m, with  variations
of about $6\%$, dependent on the covered range in $R$.
Most of this variation 
(about $5\%$) is due to the change of $\lambda _{abs}$ with
depth. The errors of the
measurement itself are estimated to be smaller than $1\%$
(i.e. 0.2\,m). We attribute the discrepancy between 
1995 and 1997 results to time variations of the water
parameters.

\begin{figure}[H]
\centering
  \mbox{\epsfig{file=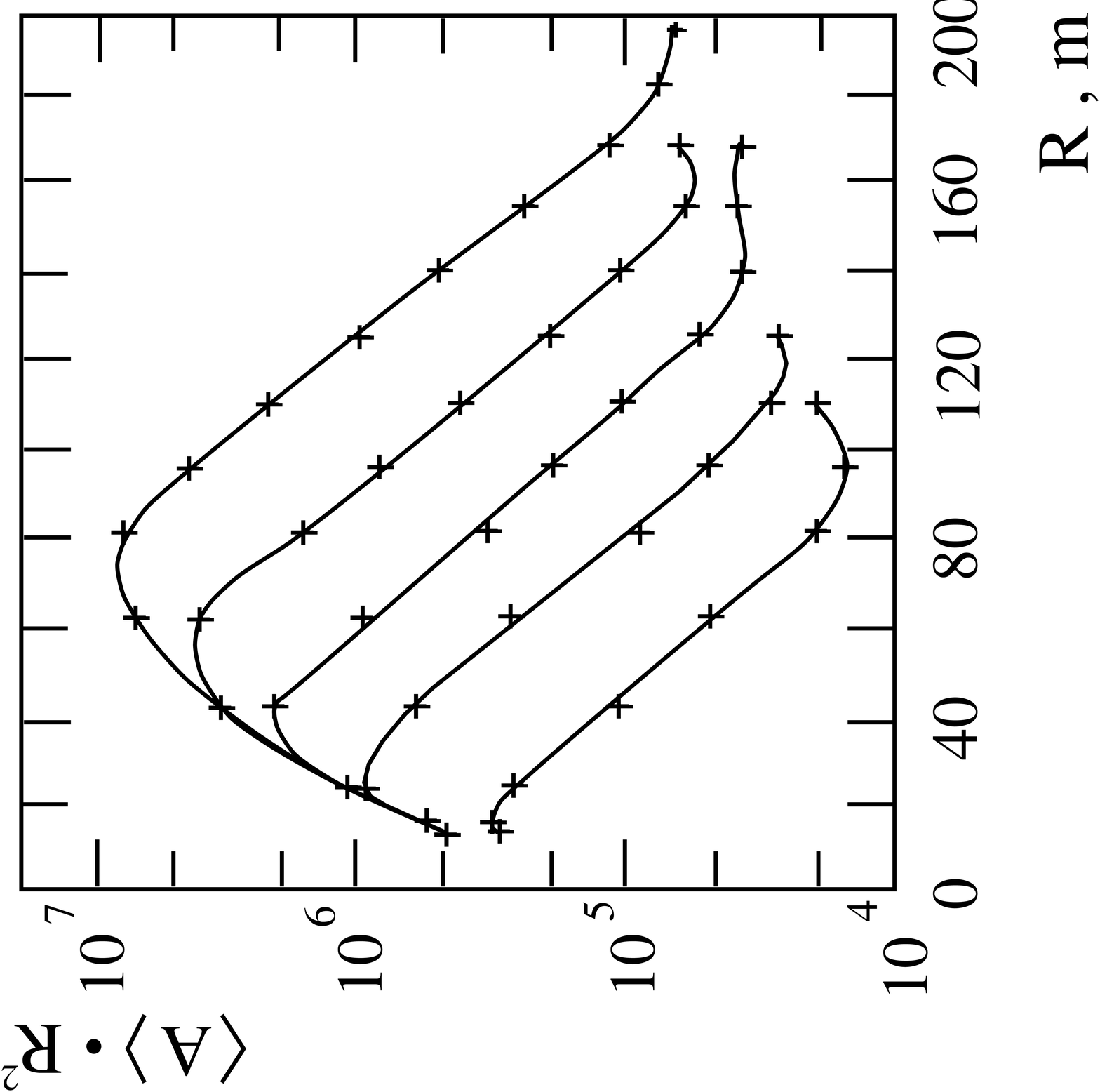,height=9cm,angle=-90}}
\caption[4]{\small
Laser experiment 1997: Logarithm of average
amplitude $\langle A \rangle$ times squared
distance $R^2$ versus distance $R$ for one of the 12 channels
of the first string. Curves are to guide the eye.
Each curve corresponds to one of the five
different laser intensities. Saturation of the curves
at low $R$ is due to the limited dynamical range, flattening
at high $R$ is due to the noise. Fits have been applied to the
linear part of the curves.
}
\end{figure}

\vspace{-3mm}
\subsection{Scattering}

To investigate scattering  we evaluated
the delay in arrival times of the photons.
According to eqs.(\ref{3.7}), (\ref{3.11}), 
these times can be written  as

\begin{equation}
\label{4.2}t=\frac R{c_w}\,\sqrt{1+\frac 13\frac R{\lambda _{eff}}}\approx
\frac 1{c_w}\,\left( R+\frac 16\frac{R^2}{\lambda _{eff}}\right) ,
\end{equation}

where $c_w$ is the velocity of light in water. 
Note that this approximation applies only for multiple scattering
under small angles. Strongly scattered photons (e.g. due to
Rayleigh scattering) arrive much later and would not coincide
with one of the many forward scattered photons. This, however,
is a condition for
a channel to trigger, since its two OMs  
are operated in coincidence.
Eq.(\ref{4.2}) does not include effects of absorption. 
Consequently, in reality the average time delay is smaller than given 
by expression (\ref{4.2}): 
photons traveling a longer path are absorbed and
contribute less to the average delay. Absorption is described by
the exponential probability distribution of eq.(14).

With $\lambda_{abs}$ determined from the intensity-vs.-distance
curve, the effect of absorption can be subtracted from the
experimental data as well as from Monte Carlo data.
Fig.9 compares the experimental delays to 
delay curves expected from different
assumptions on $\lambda_{eff}$. 
For both experiment and prediction the effect of absorption is
subtracted so that eq.(\ref{4.2}) can be applied.
The solid line gives the best fit and corresponds 
to $\lambda_{eff}=480$\,m. 
The  error of $\lambda_{eff}$ is estimated as 15\%.

\begin{figure}[H]
\centering
  \mbox{\epsfig{file=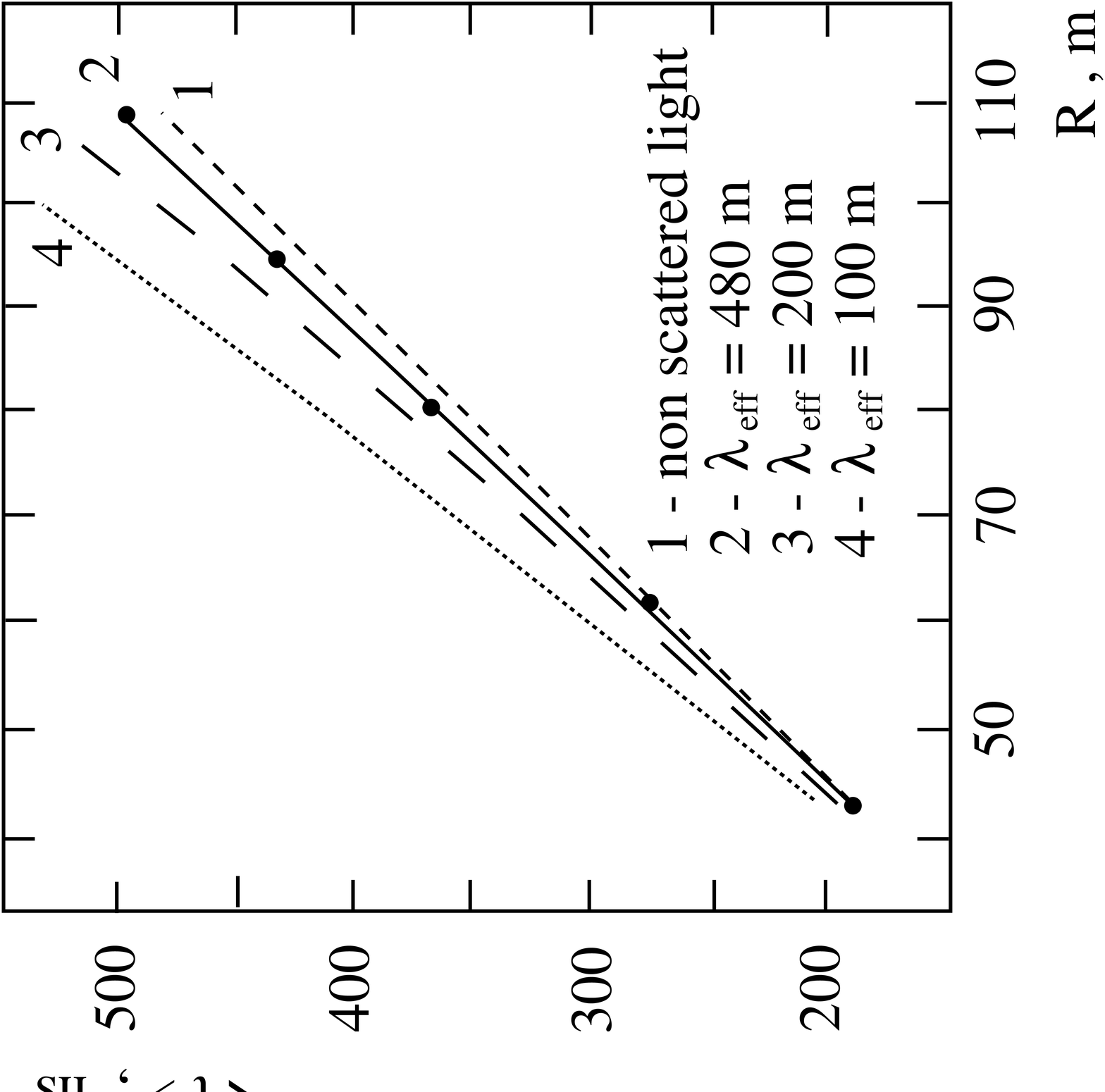,height=9cm,angle=-90}}
\caption[4]{\small
Laser experiment 1997: Average arrival times $\langle t \rangle$
vs. distance $R$.
Dots represent the measured values, lines give the expectations for
different values of the effective scattering length $\lambda_{eff}$. 
In both cases absorption effects have been subtracted (see text).
}
\end{figure}

The value of $\lambda _{eff}$ obtained in the present work can be compared
with the results of the direct measurements of $\lambda _{sct}$ and $\beta
\left( \theta \right) $ cited in \cite{Gapo96,Tara94}. There, with
the help of the measured scattering function $\beta \left( \theta \right) $,
$\left\langle \cos \left( \theta \right)
\right\rangle \approx $ 0.95\,--\,0.96 was found.
For $\lambda _{sct}$ values of 15\,--\,18\,m
have been obtained. 
In 1997 (not published), we measured 
$\left\langle \cos \left( \theta \right) \right\rangle \approx $ 0.9
and $\lambda _{sct} =$  40--50\,m.
With eq.(\ref{3.11}), 
this results in $\lambda _{eff}$ = 300\,--\,450\,m and
$\lambda _{eff}$ = 400\,--\,500\,m (1997), 
compatible
with $\lambda_{eff} = 480 \pm 70$\,m determined in the present
analysis. We note that
for $\left\langle \cos \left( \theta \right) \right\rangle $ close
to $1$ the result is extremely sensitive to the exact value of 
$\left\langle \cos \left( \theta \right) \right\rangle$.
Also, scattering parameters in Lake Baikal are known to 
vary strongly with time (see \cite{Sher}). 
This makes the agreement even more satisfactory.

\subsection{Asymptotic attenuation length}

As follows from eq.(\ref{3.14}), at large distances
\begin{equation}
\label{4.3}F\left( R\right) \sim \frac 1R\,e^{-\frac R{\lambda _{asm}}},
\end{equation}
where
\begin{equation}
\label{4.4}\lambda _{asm}=\frac{\lambda _{abs}}{1+\sqrt{\lambda
_{abs}/\lambda _{eff}}}
\end{equation}

is the so-called asymptotic attenuation length. 
Eq.(\ref{3.14}) describes the asymptotic behavior
of $F\left( R\right) $. In the asymptotic regime, 
angular and spatial dependences of $\Phi \left( R,\,\vec \Omega
,\,t\right) $ can be factorized (see, for example, \cite{Dolin}) so
that

\begin{equation}
\label{4.a}\int \,\Phi \left( R,\,\vec \Omega ,\,t\right) \,\,dt=F\left(
R\right) \rho \left( \vec \Omega \right) .
\end{equation}

Therefore, the measurement of $%
\lambda _{asm}$ is not sensitive to the orientation of the PMTs
and can be performed independently for various directions.

Although  expression (\ref{3.14}) follows from eq.(\ref{3.13}) only for
very large distances, our Monte Carlo simulations show that the 
$e^{-\frac{R}{\lambda_{asm}}}/R$ behavior is already 
seen for $R \approx 100$\,m
if $\lambda_{abs} \approx 20$\,m, $\lambda _{eff}\leq 500$\,m and
the channel faces away from the light source. Photons observed by 
these channels are preferently very often scattered and form
a light field close to the asymptotic propagation regime.
Thus the
value of $\lambda _{asm}$ can be evaluated from the data of the 1997 laser
experiment where the maximal distance $R$ was approximately 180\,m. In
fig.10 the $R$ dependence of the amplitudes for channel
$N^{\underline{0}}\,2$, looking away from the source, is shown. 
The experimental data behave exponentially 
with an exponent $1/L_{\uparrow }\approx 1/(17.6 \pm 0.5)$\,m$^{-1}$. 

\begin{figure}[H]
\centering
 \mbox{\epsfig{file=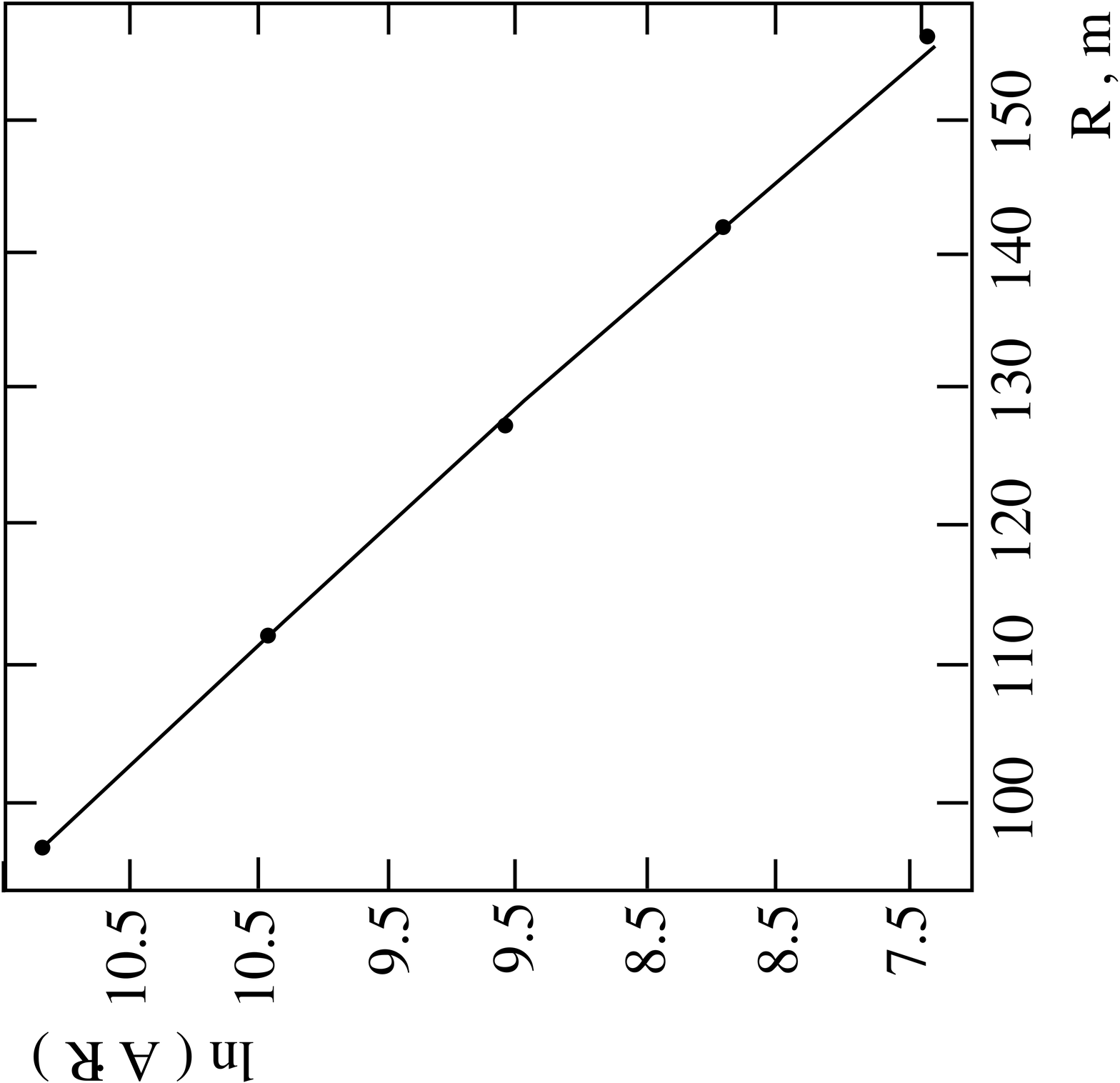,height=7.5cm,angle=-90}}
\caption[4]{\small
Logarithm of the product of average amplitude $\langle A \rangle$
and $R$ versus distance $R$ in the "asymptotic" regime
for
channel 2, looking away from the laser.
}
\end{figure}

Comparing Monte Carlo simulations to eq.(\ref{3.13}) 
we found that in the general case
$0.95 \cdot L_{\uparrow }\leq \lambda _{asm}\leq 1.05 \cdot
L_{\uparrow }$,
and so we finally obtain $\lambda
_{asm}$ = (17.6\,$\pm$\,1.5)\,m. 

Since there is a certain overlap between the regions in which both eq.(\ref
{4.3}) and eq.(\ref{3.12}) describe the photon
flux as a function of distance, 
one could try to estimate $\lambda _{eff}$ from 
$\lambda_{asm} $ and $\lambda _{abs}$, 
using eq.(\ref{4.4}). Unfortunately, the
final result is very sensitive to the exact values of asymptotic and
absorption length (cf. what was said above on the estimation of $%
\lambda _{eff} $ from $\left\langle \cos \left( \theta \right) \right\rangle
$). Therefore, the available accuracy 
of our experiments does not allow yet to
determine $\lambda_{eff}$ in this way. However, $\lambda _{asm}$ itself can
be applied as a useful parameter for the description of the
photon flux at large distances.


\vspace{-3mm}

\section{Conclusions}

\vspace{-3mm}
We have used a neutrino telescope triggered by laser light pulses
to measure the optical properties
of clear, open water with high accuracy. For 1 km deep water in Lake
Baikal, in 1997 we determined an absorption length of $22.0 \pm 1.2$\,m
and an asymptotic attenuation length of $17.6 \pm 1.5\,$m.
Quite different to standard methods, the effective scattering length 
was obtained from nanosecond timing measurements. It was measured as
$480 \pm 70$\,m which under the assumption of 
$\langle \cos \theta \rangle$ = 0.95 (0.9) 
translates to a geometrical scattering length of 24 (48)\,m.

In future, we want to refine these measurements using a larger number
of optical modules and measuring over an extended range of
distances between optical modules and laser. Also, we plan a
continuous monitoring of the optical parameters over the full year.

\bigskip
{\it This work was supported by the RFFI grant 97-05-96466.}

\end{document}